\documentclass[%
 reprint,
 superscriptaddress,
 amsmath,amssymb,
 aps,prl,
]{revtex4-2}

\usepackage{graphicx}
\usepackage{dcolumn}
\usepackage{bm}
\usepackage{hyperref}

\usepackage{xcolor}

\hypersetup{colorlinks=true, allcolors=blue}

\begin{document}

\preprint{APS/123-QED}

\title{Collisionless ion-electron energy exchange in magnetized shocks}

\author{Y. Zhang} \email{vyzhang@lle.rochester.edu}
    \affiliation{Laboratory for Laser Energetics, University of Rochester, Rochester, New York 14623, USA}
    \affiliation{Department of Mechanical Engineering, University of Rochester, Rochester, New York 14627, USA}
\author{P. V. Heuer}
    \affiliation{Laboratory for Laser Energetics, University of Rochester, Rochester, New York 14623, USA}
\author{H. Wen}
    \affiliation{Laboratory for Laser Energetics, University of Rochester, Rochester, New York 14623, USA}
\author{J. R. Davies}
    \affiliation{Laboratory for Laser Energetics, University of Rochester, Rochester, New York 14623, USA}
    \affiliation{Department of Mechanical Engineering, University of Rochester, Rochester, New York 14627, USA}
\author{C. Ren}
    \affiliation{Laboratory for Laser Energetics, University of Rochester, Rochester, New York 14623, USA}
    \affiliation{Department of Mechanical Engineering, University of Rochester, Rochester, New York 14627, USA}    
    \affiliation{Department of Physics and Astronomy, University of Rochester, Rochester, New York 14627, USA}
\author{D. B. Schaeffer}
    \affiliation{Department of Physics and Astronomy, University of California-Los Angeles, Los Angeles, California 90095, USA}
\author{E. G. Blackman}
    \affiliation{Laboratory for Laser Energetics, University of Rochester, Rochester, New York 14623, USA}
    \affiliation{Department of Physics and Astronomy, University of Rochester, Rochester, New York 14627, USA}
\author{J. Zhang}
    \affiliation{Laboratory for Laser Energetics, University of Rochester, Rochester, New York 14623, USA}
    \affiliation{The Institute of Optics, University of Rochester, Rochester, New York 14627, USA}
\author{A. Bret}
    \affiliation{ETSI Industriales, Universidad de Castilla-La Mancha, 13071 Ciudad Real, Spain}
    \affiliation{Instituto de Investigaciones Energéticas y Aplicaciones Industriales, Campus Universitario de Ciudad Real, 13071 Ciudad Real, Spain}
\author{F. García-Rubio}
    \affiliation{Pacific Fusion Corporation, Fremont, California 94538, USA}

\date{\today}

\begin{abstract}
Energy partition between ions and electrons in collisionless shocks has long been an unsolved fundamental physical problem. We show that kinetic simulations of moderate Alfvénic Mach number, magnetized, collisionless shocks reveal rapid, faster-than-Coulomb, energy exchange between ions and electrons when the plasma is sufficiently magnetized. Using kinetic and multi-fluid models with counter-streaming ions, we identify resonances between electron whistler and ion magnetohydrodynamic waves that account for this rapid energy exchange.
\end{abstract}


\maketitle


Magnetized collisionless shocks, such as planetary shocks, supernova remnants, and galaxy clusters, are ubiquitous in the Universe, governing energy dissipation and particle acceleration. A key open question is how energy is partitioned between particle species across such shocks. Coulomb collisions are rare in the aforementioned astrophysical environments due to the low plasma densities and large mean free paths \cite{spitzer1941dynamics}. The low collisionality is at odds with observations that electron-to-ion temperature ratios are on the order of unity, $T_e/T_i\sim$ 0.03 - 2 \cite{schwartz1988electron,ghavamian2013electron,schwartz2022energy}, which necessitates an effective ion-electron energy exchange.

Lasers allow the creation of analogous environments in the laboratory. The formation of Weibel-mediated collisionless shocks and particle acceleration in an initially unmagnetized plasma has been extensively reported \cite{sakawa2024laser,fiuza2012laser,zhang2017collisionless,yao2023high,fiuza2020electron,yao2021laboratory,li2019collisionless,ross2013collisionless}. Recent experimental developments have enabled the study of magnetized shocks \cite{schaeffer2017generation,schaeffer2019direct,endrizzi2021laboratory,dover2025optical,tang2025laboratory}, bridging the gap between laboratory and astrophysical conditions. Anomalous electron heating is observed in these experiments, especially in magnetized cases.

Numerical and theoretical studies have significantly advanced our comprehension of magnetized collisionless shocks \cite{silva2004proton,lemoine2019physics,bohdan2021magnetic,gong2023electron,vanthieghem2024electron,ghizzo2023collisionless,orusa2023fast,marret2022enhancement}. Combined kinetic and test particle simulations \cite{vanthieghem2024electron} have indicated that an interplay between an ambipolar electric field and Weibel-mediated turbulence is responsible for electron heating in weakly magnetized, high Mach number shocks. Kinetic simulations by Liu \cite{liu2024ion} confirm the importance of perpendicular magnetization, which significantly amplifies the field energy and modifies particle distributions.

In this Letter, we report particle-in-cell (PIC) simulations where ions and electrons are treated kinetically. To mimic the transition region of a collisionless shock, where reflected ions co-exist with incoming ions \cite{zhang2024kinetic}, we simulate two counter-streaming ion flows and a thermal electron background. We focus on (quasi-)parallel shocks where the angle between flow and imposed $B$-field is less than $45^\circ$, which are more effective at heating electrons \cite{zhang2025magnetized}, using moderate Alfvénic Mach numbers ($M_A \lesssim 20$), relevant to Earth’s bow shock \cite{lalti2022database} and accessible in laboratory experiments. The energy exchange observed in self-consistent shock simulations \cite{zhang2024kinetic} is well-reproduced by our present simulations. Using a multi-fluid framework, we identify multiple resonances between electron whistler (EW) and ion magnetohydrodynamic (MHD) waves, which account for the energy exchange, and may seed turbulence. Although quasi-parallel shocks are difficult to create in the laboratory \cite{zhang2021kinetic,zhang2021kinetic2,heuer2020laser,zhang2024kinetic}, counter-streaming flows achievable on laser systems such as OMEGA can be used to validate the proposed mechanism, complementary to satellite data \cite{lalti2022database,li2025direct} of shocks that are typically fully formed.

Our simulations are performed with OSIRIS 4, a fully kinetic PIC code \cite{fonseca2002osiris}. Our 2-D simulations utilize periodic boundary conditions for particles and fields in both directions. We set up two counter-streaming ion species with drifting Maxwellian distributions in the electron rest frame. The drift velocities obey the zero-current condition, $n_{i1}M_{A,i1} + n_{i2}M_{A,i2} = 0$, where $n$ is density, the subscripts $i1$ ($i2$) and $e$ denote ion 1 (ion 2) and electron populations, respectively. The angle between the ion drifts and the external in-plane $B$-field is $\alpha$. An ion mass $m_i = 400 m_e$ is used. The initial temperatures of all species are set to $T$ = 100 eV ($\beta\equiv\sum_{s=i1,i2,e}n_sT_s/B^2=1.12$). We utilize a 1280 × 1280 simulation box with a grid size of 0.5 $c/\omega_{pe}$, and time step of 0.35 $\omega_{pe}^{-1}$ to ensure Courant convergence, where $\omega_{pe}$ is the plasma frequency, and $c$ is the speed of light. The collision module in OSIRIS is turned off. Numerical collisions are kept low by using a minimum of 10 × 10 macro-particles for each species per cell. Total energy is conserved to within 0.3\%. 

For moderate Mach number, quasi-parallel shocks, 10-20\% of upstream ions are reflected and form a shock transition region where shock-reflected ions interact with upstream ions in an electron background \cite{wilson2016relativistic}. The width of transition region in quasi-parallel shocks is $\sim$100 ion inertial lengths \cite{zhang2024kinetic}. We first simulate shock-transition relevant parameters with $n_{i1}/n_{i2}$ = 0.85/0.15; $M_{A,i1}$, $M_{A,i2}$ = 1.875, $-10.625$; and $\alpha = 0^\circ$ (labeled S: $\alpha = 0^\circ$). The total energy of a species is divided into bulk energy $E_b = \frac{1}{2}m\langle v \rangle^2$ and thermal energy $E_{th} = \langle E \rangle - E_b \equiv \frac{3}{2} \cdot T$, where $\langle x \rangle \equiv \int xf(x)dx$ represents the average of a normalized distribution $f(x)$, and $T$ denotes the effective temperature. 

\begin{figure}
    \centering
    \includegraphics[width=1\linewidth]{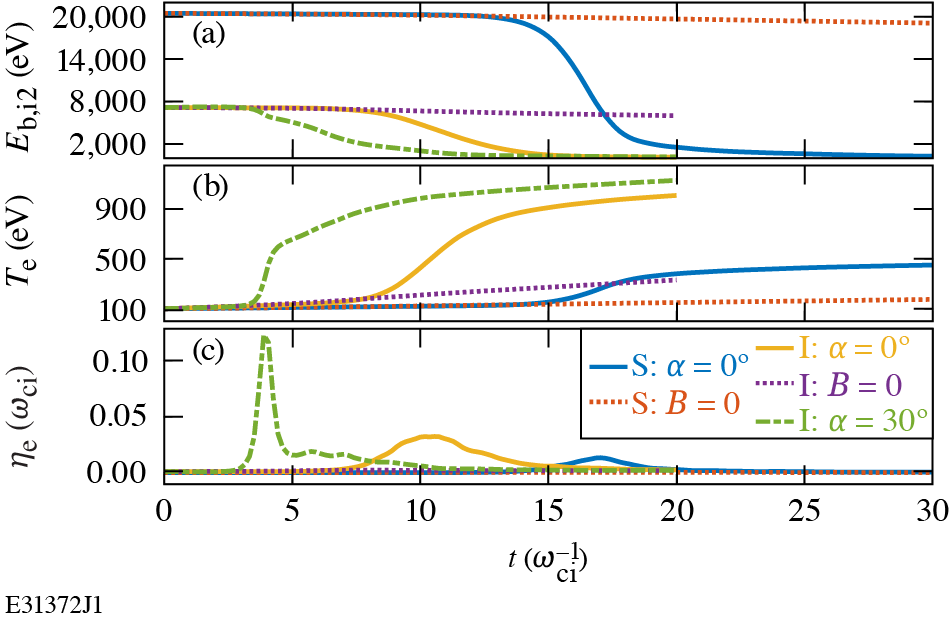}
    \caption{Time history of (a) ion bulk energy, (b) effective electron temperature, and (c) electron heating rate (defined in the text) in two types of simulations: S represents shock transition-region related parameters, I refers to instability simulations. Time for $B = 0$ is normalized by the same factor as $B\neq0$.}
    \label{fig:E-t}
\end{figure}

Figures~\ref{fig:E-t}(a) and (b) show the time evolution of $E_{b,i2}$ and $T_e$, respectively. $E_{b,i1}$ is omitted because the majority of the bulk energy is in $E_{b,i2}$ due to the larger drift velocity. In Fig.~\ref{fig:E-t}(a), the ion bulk energy almost depletes after 25 ion gyro-times [$\omega_{ci}^{-1}\equiv(eB/m_ic)^{-1}$], after which, without further effective drive from the ion bulk motion, the system reaches a slowly-evolving stage. Figure 1(b) shows that the electrons are heated on a timescale of $\sim10\omega_{ci}^{-1}$. We define an electron heating rate $\eta_e \equiv (dE_{th,e}/dt)/|E_{th,e}-E_{bal}|$, where $E_{bal} = \frac{3}{2} \cdot T_{bal}$, $bal$ denoting the balanced state of a fully thermalized system (maximum entropy). The electron heating and ion-electron energy partition observed here are consistent with PIC simulations of magnetized shocks \cite{zhang2024kinetic} and astronomical observations \cite{schwartz1988electron,ghavamian2013electron,schwartz2022energy}.

The importance of magnetization is examined by turning off the applied $B$-field in the simulations (S: $B$ = 0). The change of the ion bulk energy [Fig.~\ref{fig:E-t}(a)], electron temperature [Fig.~\ref{fig:E-t}(b)], and electron heating rate [Fig.~\ref{fig:E-t}(c)] over the simulation time are significantly smaller than the magnetized simulations. Magnetization is therefore proven to be a necessary condition for an effective energy exchange.

\begin{figure}
    \centering
    \includegraphics[width=1\linewidth]{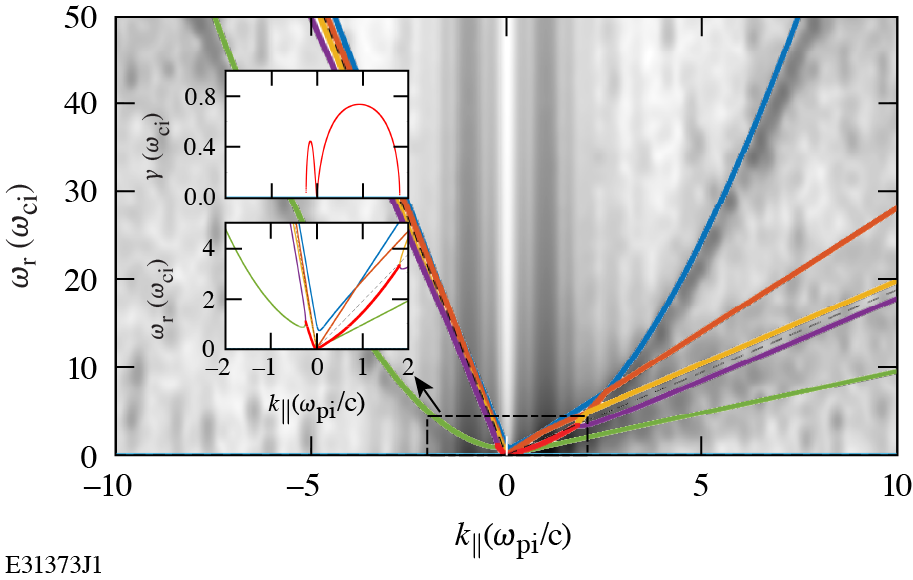}
    \caption{Fourier spectrum (greyscale) from the magnetized simulations in the early-stage 0 – $5\omega_{ci}^{-1}$. The lines are the modes (real frequency) from the multi-fluid dispersion relation, the parts in red indicate unstable modes. The upper and lower insets show the growth rate of the unstable modes and a magnified view of the real frequency, respectively.}
    \label{fig:w-k}
\end{figure}

One possible mechanism is that the magnetic field is required because the energy exchange involves the interaction of magnetized plasma waves. Figure 2 shows the Fourier spectrum in real frequency-parallel (to \textbf{B}) wavenumber ($\omega_r$-$k_{\parallel}$) space from the early-stage of the magnetized simulations (0 – $5\omega_{ci}^{-1}$), where several modes are observed, including a parabola-shaped one, and two modes at lower frequencies and wavenumbers. A multi-fluid dispersion relation solver \cite{supp} was developed to identify instabilities. In Fig.~\ref{fig:w-k}, the modes from the solver are overlaid. Unstable modes are indicated in red, with growth rates, $\gamma$'s, shown in the upper inset. Good correspondence is seen between the simulations and theoretical results. The parabola-shaped shadows are identified as EW modes. Most importantly, the unstable modes that the solver predicts in $-0.25$ - 0 $\omega_{pi}/c$ and 0 - 1.8 $\omega_{pi}/c$ are observed in the spectrum. These modes are not seen in the unmagnetized simulations, where no efficient energy exchange process is observed.

To facilitate the identification of the instability, we use a field-parallel ($\alpha = 0^\circ$) setup with equal-density ions ($n_{i1} = n_{i2}$) and Alfvénic Mach numbers $M_{A,i1}$, $M_{A,i2}$ = $\pm6.25$. These symmetrical simulations are labeled I. The evolution of $E_{b,i2}$, $T_e$, and $\eta_e$ are shown in Fig.~\ref{fig:E-t}. Similar to S simulations, efficient energy exchange is observed on a timescale $\lesssim20\omega_{ci}^{-1}$ in the I: $\alpha = 0^\circ$ simulations. Turning off the external $B$-field (I: $B$ = 0) weakens the process significantly. The I: $\alpha = 30^\circ$ simulations show an earlier and faster process than I: $\alpha = 0^\circ$, but with a similar slowly-evolving stage. This is consistent with the observations in the shock simulations that quasi-parallel shocks form faster than parallel shocks \cite{zhang2024kinetic}.

\begin{figure}
    \centering
    \includegraphics[width=1\linewidth]{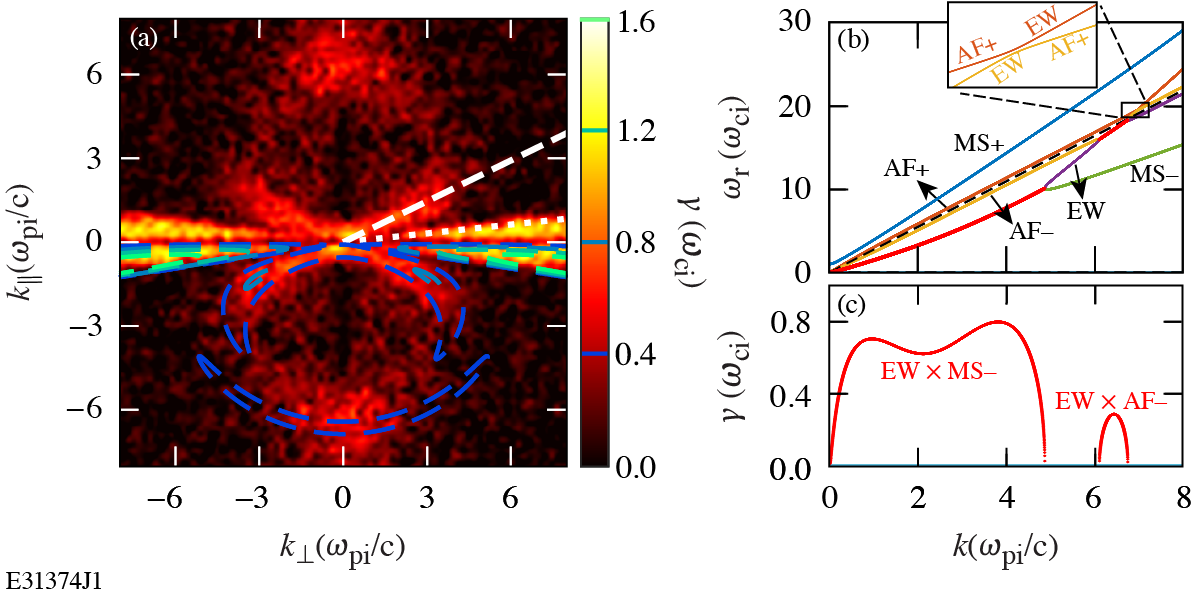}
    \caption{(a) Growth rates inferred from the out-of-plane $B$-field for the symmetrical, parallel simulations (I: $\alpha = 0^\circ$). Contours in the lower half show the solutions from the multi-fluid dispersion relation. Dispersion relation at $\theta_k = 64^\circ$: (b) real frequency and (c) growth rate. The red portions of the branches are unstable.  EW = electron whistler, AF = Alfvén, MS = magnetosonic, +/$-$ = forward/backward, A$\times$B denotes a resonance between modes A and B.}
    \label{fig:DR+64deg}
\end{figure}

Figure~\ref{fig:DR+64deg}(a) shows the linear growth rate in $k$-space, inferred from the out-of-plane $B$-field in the I: $\alpha = 0^\circ$ simulations over 0 – $4\omega_{ci}^{-1}$. Modes with $\gamma\sim\omega_{ci}$ are observed, including modes over a range of $\theta_k$ (angle between the $k$-vector and applied $B$-field), and a nearly-perpendicular mode. Contours of the theoretical growth rates are overlaid in the lower half. The theoretical results capture the main structures and growth rates measured in the PIC simulations. Figures~\ref{fig:DR+64deg}(b) and (c) give the multi-fluid dispersion curves at $\theta_k = 64^\circ$, marked by the white dashed line in Fig.~\ref{fig:DR+64deg}(a). In Fig.~\ref{fig:DR+64deg}(b), each curve on the $\omega_r$-$k$ plane represents a wave mode. Fig.~\ref{fig:DR+64deg}(c) shows the corresponding $\gamma$'s. 

From the real frequencies where the curves do not merge in Fig.~\ref{fig:DR+64deg}(b), a group of low-frequency waves are identified, including EW waves, and ion MHD waves: Alfvén (AF) and magnetosonic (MS) waves. Note that ion MHD waves co-drift with the ion bulk motions at $\omega/k = V_d\cos\theta_k$, marked by the black dashed line in Fig.~\ref{fig:DR+64deg}(b). Each of the MS and AF waves has a forward and backward wave. The forward (backward) wave, marked by + ($-$), travels faster (slower) than the drift. We thus have 9 low-frequency waves: 4 ion MHD waves (MS$\pm$, AF$\pm$) from each ion drift, and EW waves. Figure~\ref{fig:DR+64deg}(b) shows the wave modes that have positive propagation ($\omega_r/k\geq0$), negative propagation ($\omega_r/k\leq0$) is omitted due to the symmetry, thus only MHD waves in the ion 1 species and EW waves are plotted. Such EW and MHD waves are frequently detected in the planetary magnetosphere and interplanetary space \cite{jagarlamudi2021whistler,karbashewski2023whistler,huang2020excitation,kumar2024direct}, and are known to influence energy exchange with electrons \cite{kitamura2022direct,zhong2022evidence,li2025direct}.

The nonzero $\gamma$'s in Fig.~\ref{fig:DR+64deg}(c) correspond to the mode mergers in Fig.~\ref{fig:DR+64deg}(b), highlighted in red. Two dispersion curves passing close together can allow resonant exchange of energy between the modes. At low frequencies, EW waves can interact with ion MHD waves. The frequency match means that the mechanism is a resonance. At $\theta_k = 64^\circ$, the unstable mode at 0 – $5\omega_{pi}/c$ is the resonance between EW and MS$-$ (EW×MS$-$) while the mode at 6 – $7\omega_{pi}/c$ is EW×AF$-$. The resonance also operates at other $\theta_k$’s, constituting two main modes in Fig.~\ref{fig:DR+64deg}(a). The inner part is EW×MS$-$, and the outer is EW×AF$-$.

However, in Fig.~\ref{fig:DR+64deg}(b), not every curve intersection leads to growth. For example, EW does not resonate with AF+ in this setup around $k\approx7\omega_{pi}/c$. This is because the polarization of the waves must also match, so that angular momentum conservation allows the transfer of energy between the waves. The polarization match is discussed in the supplemental material \cite{supp}. The inset in Fig.~\ref{fig:DR+64deg}(b) shows that EW and AF+ waves switch near $k\approx7\omega_{pi}/c$ without a true crossing, which is a mode transformation \cite{keppens2019fresh}. 

MS waves do not only interact with EW waves. Ion-ion resonance can also be activated at nearly-perpendicular angles. Using the same method above, the mode near $\theta_k = 84^\circ$, marked by the white dotted line in Fig.~\ref{fig:DR+64deg}(a), is identified as the ion-ion coupling MS$-$(i1)×MS$-$(i2) \cite{supp}. However, the unstable ion-ion MS mode does not contribute to energy exchange between ions and electrons. In the unmagnetized case (I: $B$ = 0), acoustic modes also become unstable at perpendicular angles \cite{supp} yet no effective energy exchange is activated. Therefore, the ion-ion magnetosonic (acoustic in the unmagnetized case) mode does not drive a significant ion-electron energy exchange. 

\begin{figure}
    \centering
    \includegraphics[width=1\linewidth]{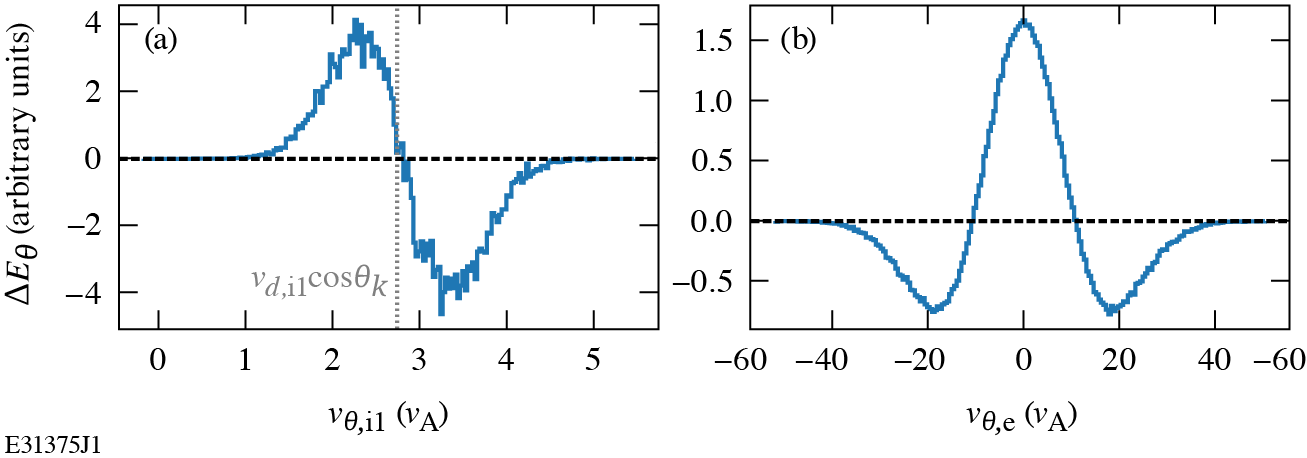}
    \caption{Energy gain, $\Delta E_\theta$, as a function of (a) ion 1 and (b) electron velocities, $v_\theta$, in the $\theta_k = 64^\circ$ direction. Positive (negative) $\Delta E_\theta$ indicates particles gaining (losing) energy.}
    \label{fig:landau}
\end{figure}

Further study of the PIC results shows that the energy exchange is achieved via wave-particle interactions (WPIs) \cite{gendrin1983wave}. In WPIs, the waves represent an efficient conduit for transferring energy and the wave energy is usually several orders of magnitude smaller than the kinetic energy of particles \cite{shklyar2017energy}. Figure~\ref{fig:landau}(a) shows the energy gain in the $\theta_k = 64^\circ$ direction, $\Delta E_\theta$, as a function of the initial velocity in the same direction, $v_\theta$, of ion 1 species over 0 – $0.2\omega_{ci}^{-1}$. Ions with larger (smaller) velocities lose (gain) energy. The separation velocity matches the phase velocity of the unstable waves $V_{d,i1}\cos\theta_k = 2.7v_A$, which is a signature of Landau resonance \cite{landau1946vibrations}. Figure~\ref{fig:landau}(b) shows a similar relation for electrons. Although electron velocities span a larger range, the observation that electrons with larger (smaller) velocities lose (gain) energy holds. The ion and electron heating are therefore achieved by Landau resonance.

Landau resonance does not necessarily operate at the same rate among $\theta_k$. The simulations show an almost isotropic electron heating over ion timescales \cite{supp}. This is achieved through the help of magnetic mirroring and anisotropy-driven instabilities \cite{gary2006linear}. These effects isotropize the electrons over the timescale of hundreds of $\omega_{ce}^{-1}$. Therefore, on the timescale of tens of $\omega_{ci}^{-1}$ ($\gg\omega_{ce}^{-1}$), nearly isotropic electron heating is observed.

Hence, the energy exchange is a two-stage process: 1) the unstable drifting ion velocity distribution functions are the energy source, through wave resonances (EW$\times$MHD waves), energy is transferred to the unstable waves (resonant modes); 2) via Landau resonance, the waves serve as a mediator, transferring the energy and thermalizing ions and electrons. The process works comparably with a realistic mass ratio \cite{supp}.

\begin{figure}
    \centering
    \includegraphics[width=1\linewidth]{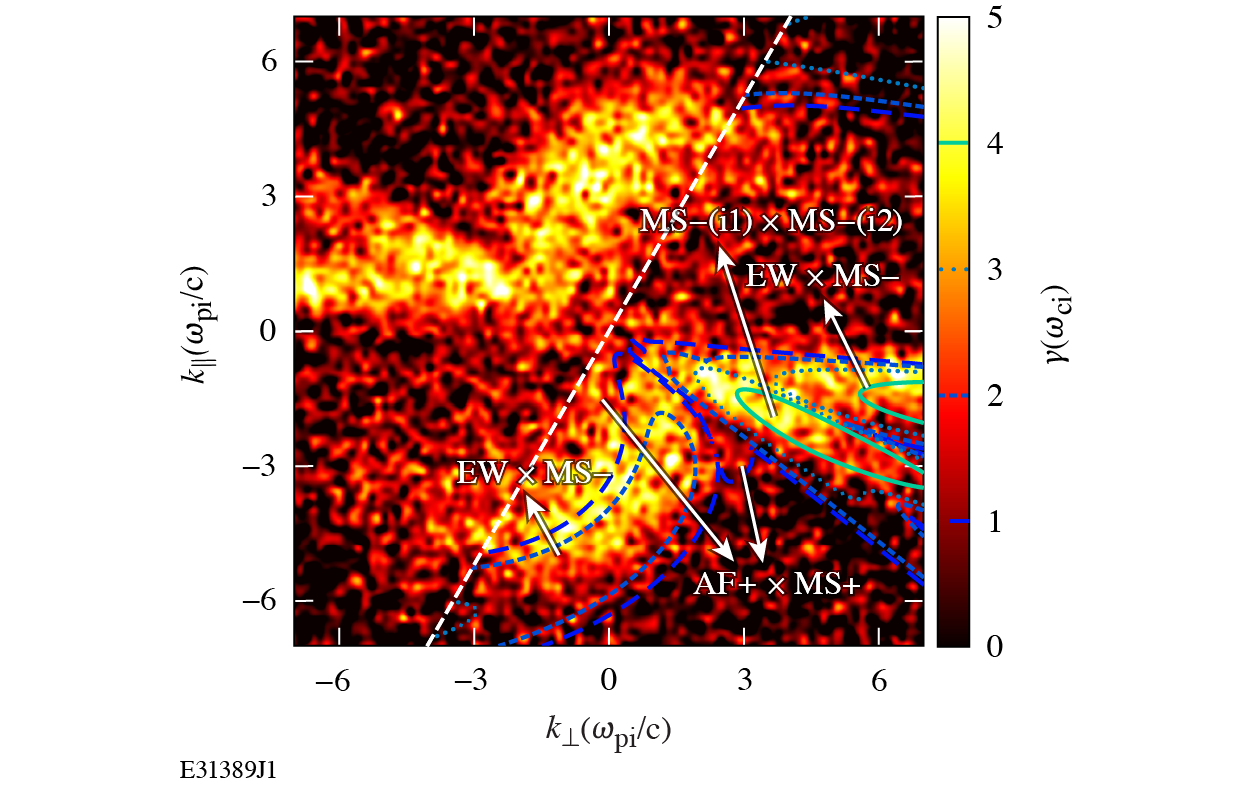}
    \caption{Growth rates inferred from the out-of-plane $B$-field for the I: $\alpha = 30^\circ$ simulations. Contours on the right half show the solutions from the multi-fluid dispersion relation. The dashed white line is the ion drift direction.}
    \label{fig:alpha-30deg}
\end{figure}

Similar, albeit more complicated results, are found in the oblique-field case. Fig.~\ref{fig:alpha-30deg} shows the growth rates inferred over 0 – $1\omega_{ci}^{-1}$ for the I: $\alpha = 30^\circ$ simulations. Many modes are observed with $\gamma\lesssim 5\omega_{ci}$, larger than I: $\alpha = 0^\circ$, which suggests a faster energy exchange process, in agreement with Fig.~\ref{fig:E-t} and shock simulations \cite{zhang2024kinetic}. Solutions of the dispersion relation are overlaid as contours on the right half. The main structures and growth rates agree well with the simulations.

The identified modes include EW$\times$MS$-$, AF+$\times$MS+, and MS$-$(i1)$\times$MS$-$(i2), labeled in Fig.~\ref{fig:alpha-30deg}, and several weaker ones (unlabeled). In the oblique setup, modes overlap in $k$-space and become very complicated. The comparison to the simulation result suggests that the most prominent resonances are EW$\times$MS$-$ and MS$-$(i1)$\times$MS$-$(i2).

The proposed wave resonances are extensions to previously studied instabilities. Classic right-hand instability (RHI) and non-resonant instability (NRI, or Bell instability \cite{bell2004turbulent}), which have been known to be responsible for parallel shock formation \cite{heuer2018observations}, can be recovered from the multi-fluid model discussed above with $\alpha = 0^\circ$, $\theta_k = 0^\circ$, and $T = 0$. RHI is identical to EW$\times$AF$-$(i1) propagating in the positive direction, and NRI is EW$\times$AF$-$(i2) propagating negatively. Yet in a warm plasma with a finite temperature \cite{weidl2019three}, at $\theta_k = 0^\circ$, NRI and RHI are replaced by EW$\times$MS$-$(i1) and EW$\times$MS$-$(i2), respectively, because of the non-zero thermal pressure. Now we can show that, in Fig.~\ref{fig:w-k}, the unstable modes near the origin are positive-propagating EW$\times$MS$-$(i1) and negative-propagating EW$\times$MS$-$(i2). 

In conclusion, magnetized, collisionless ion-electron energy exchange has been studied with 2-D kinetic simulations in the quasi-parallel, moderate Alfvénic Mach number regime. A simplified magnetized counter-streaming multi-species setup reproduces the faster-than-Coulomb collisionless energy exchange process and ion-electron energy partition observed in collisionless astronomical systems \cite{schwartz1988electron,ghavamian2013electron,schwartz2022energy}, and shock simulations \cite{zhang2024kinetic}. We have shown that magnetization is necessary for an effective energy exchange. Calculations from a multi-fluid framework agree with the growing modes observed in the simulations. The modes are identified as the resonances between EW waves and drifting ion MHD waves. The ion drifts drive the instability via the wave resonances, then the energy is transferred from the instability to thermal particles via Landau resonance. An oblique setup produces a more complicated spectrum of waves, but the proposed resonances from the multi-fluid model still agree well with the simulations. Modes with $\gamma\sim\omega_{ci}$ can facilitate substantial ion-electron energy exchange, bringing $T_e/T_i$ to order unity after several $\omega_{ci}^{-1}$'s, or $(V_{sh}/\omega_{ci})$'s behind the shock, where $V_{sh}$ is the shock velocity. These findings could potentially be implemented into two-temperature MHD or hybrid simulation frameworks as a proxy for ion-electron energy exchange.

Our demonstration that faster-than-Coulomb coupling between ions and electrons in collisionless shocks is greatly enhanced by wave resonances that depend on a magnetic field raises the question as whether a similar conclusion also arises for subsonic turbulent collisionless astrophysical plasmas as well. Evidence from magnetic reconnection simulations provides other evidence along these lines \cite{zhang2023ion}.
Identification of faster-than-Coulomb couplings generally expands the utility of the MHD approximation for more weakly collisionless regimes, as  found empirically in recent simulations \cite{achikanath2024critical}. On the other hand, two-temperature accretion flow models in magnetized plasmas that depend only  on the Coulomb coupling rate for energy transfer to electrons \cite{yuan2014hot} would be constrained.

\begin{acknowledgments}
We thank Dr. Lynn Wilson for helpful discussions. This material is based upon work supported by the Department of Energy (National Nuclear Security Administration) University of Rochester “National Inertial Confinement Fusion Program” under Award No. DE-NA0004144, the Department of Energy under Award Nos. DE-SC0020431 and DE-SC0024566, and the resources of the National Energy Research Scientific Computing Center (NERSC), a U.S. Department of Energy Office of Science User Facility located at Lawrence Berkeley National Laboratory. A.B. acknowledges support by the Ministerio de Economía y Competitividad of Spain (Grant No. PID2021-125550OBI00). The authors thank the UCLA-IST OSIRIS consortium for the use of OSIRIS. This report was prepared as an account of work sponsored by an agency of the U.S. Government. Neither the U.S. Government nor any agency thereof, nor any of their employees, makes any warranty, express or implied, or assumes any legal liability or responsibility for the accuracy, completeness, or usefulness of any information, apparatus, product, or process disclosed, or represents that its use would not infringe privately owned rights. Reference herein to any specific commercial product, process, or service by trade name, trademark, manufacturer, or otherwise does not necessarily constitute or imply its endorsement, recommendation, or favoring by the U.S. Government or any agency thereof. The views and opinions of authors expressed herein do not necessarily state or reflect those of the U.S. Government or any agency thereof.
\end{acknowledgments}


\bibliographystyle{apsrev4-2}
\bibliography{Resonance}

\end{document}


\preprint{APS/123-QED}

\title{Collisionless ion-electron energy exchange in magnetized shocks\\
Supplemental Material}

\author{Y. Zhang} \email{vyzhang@lle.rochester.edu}
    \affiliation{Laboratory for Laser Energetics, University of Rochester, Rochester, New York 14623, USA}
    \affiliation{Department of Mechanical Engineering, University of Rochester, Rochester, New York 14627, USA}
\author{P. V. Heuer}
    \affiliation{Laboratory for Laser Energetics, University of Rochester, Rochester, New York 14623, USA}
\author{H. Wen}
    \affiliation{Laboratory for Laser Energetics, University of Rochester, Rochester, New York 14623, USA}
\author{J. R. Davies}
    \affiliation{Laboratory for Laser Energetics, University of Rochester, Rochester, New York 14623, USA}
    \affiliation{Department of Mechanical Engineering, University of Rochester, Rochester, New York 14627, USA}
\author{C. Ren}
    \affiliation{Laboratory for Laser Energetics, University of Rochester, Rochester, New York 14623, USA}
    \affiliation{Department of Mechanical Engineering, University of Rochester, Rochester, New York 14627, USA}    
    \affiliation{Department of Physics and Astronomy, University of Rochester, Rochester, New York 14627, USA}
\author{D. B. Schaeffer}
    \affiliation{Department of Physics and Astronomy, University of California-Los Angeles, Los Angeles, California 90095, USA}
\author{E. G. Blackman}
    \affiliation{Laboratory for Laser Energetics, University of Rochester, Rochester, New York 14623, USA}
    \affiliation{Department of Physics and Astronomy, University of Rochester, Rochester, New York 14627, USA}
\author{J. Zhang}
    \affiliation{Laboratory for Laser Energetics, University of Rochester, Rochester, New York 14623, USA}
    \affiliation{The Institute of Optics, University of Rochester, Rochester, New York 14627, USA}
\author{A. Bret}
    \affiliation{ETSI Industriales, Universidad de Castilla-La Mancha, 13071 Ciudad Real, Spain}
    \affiliation{Instituto de Investigaciones Energéticas y Aplicaciones Industriales, Campus Universitario de Ciudad Real, 13071 Ciudad Real, Spain}
\author{F. García-Rubio}
    \affiliation{Pacific Fusion Corporation, Fremont, California 94538, USA}

\date{\today}

\maketitle

\onecolumngrid

\section{Multi-fluid dispersion relation solver}
The dispersion relation solver considers a multi-species configuration, in which each species is modeled by a single fluid. The system is governed by mass and momentum conservation and Maxwell's equations. Density $n$, velocity \textbf{v}, electric field \textbf{E}, and magnetic field \textbf{B} are divided into equilibrium and perturbed quantities, $f = f_0 + f_1\exp(i\textbf{k}\cdot\textbf{r}-i\omega t)$, where $f_0 \gg f_1$. Without an external $E$-field ($\textbf{E}_0 = 0$), the first order equations are
\begin{equation}\begin{aligned}
{\frac{\partial n_1}{\partial t}} + n_0\nabla\cdot\textbf{v}_1 + \textbf{v}_0\cdot\nabla n_1 = 0,
\label{eq:con-mass}\end{aligned}\end{equation}
\begin{equation}\begin{aligned}
mn_0\lbrack{\frac{\partial \textbf{v}_1}{\partial t}} + (\textbf{v}_0\cdot\nabla)\textbf{v}_1\rbrack=
qn_0(\textbf{E}_1+\frac{\textbf{v}_1\times\textbf{B}_0}{c}+\frac{\textbf{v}_0\times\textbf{B}_1}{c})
+qn_1(\frac{\textbf{v}_0\times\textbf{B}_0}{c})-\Gamma k_BT\nabla n_1,
\label{eq:con-momentum}\end{aligned}\end{equation}
\begin{equation}\begin{aligned}
\nabla\times\textbf{E}_1=-\frac{1}{c}\frac{\partial\textbf{B}_1}{\partial t},
\label{eq:faraday}\end{aligned}\end{equation}
\begin{equation}\begin{aligned}
\nabla\times\textbf{B}_1=\frac{4\pi}{c}\textbf{J}_1+\frac{1}{c}\frac{\partial\textbf{E}_1}{\partial t},
\label{eq:ampere}\end{aligned}\end{equation}
where $\Gamma = 5/3$ is the polytropic index, $q$ is the charge, $c$ is the speed of light, $\textbf{J}_1=\sum_s q_s (n_{0,s}\textbf{v}_{1,s}+n_{1,s}\textbf{v}_{0,s})$ is the current. The polytropic equation of state in Eq. (\ref{eq:con-momentum}) completes the system of equations. 

In Cartesian coordinates, the conservation of momentum [Eq. (\ref{eq:con-momentum})] has 3 equations in 3 directions for each species. Maxwell’s equations Eq. (\ref{eq:faraday}) and (\ref{eq:ampere}) are expanded to 6 equations. Therefore, we have a system of ($4N_s+6$) equations, where $N_s$ is the number of species. The linearized equations can be numerically solved at any real wavenumber $k$ by implementing a complex $\omega = \omega_r + \gamma i$, where $\omega_r$ is the real frequency and $\gamma$ is the growth rate. In our case, we have 3 species (ion 1, ion 2, and electron), hence, the dispersion relations are solved from 18 linearized equations. 

\section{Polarization match condition for the wave resonance}
We consider a wave propagating in the in the $+\textbf{z}’$ direction, i.e., $\textbf{k}\parallel \textbf{z}’$. The wave polarization state is defined as $P_{k(z’)}\equiv iB_{x’}/B_{y’}$, where a positive (negative) $P_k$  corresponds to a right-hand (left-hand) polarized wave. In particular, $P_k =\pm1$ denotes a right-hand or left-hand circularly polarized (RCP/LCP) wave, respectively \cite{fitzpatrick2022plasma}.

\begin{figure}[H]
    \centering
    \includegraphics[width=0.4\linewidth]{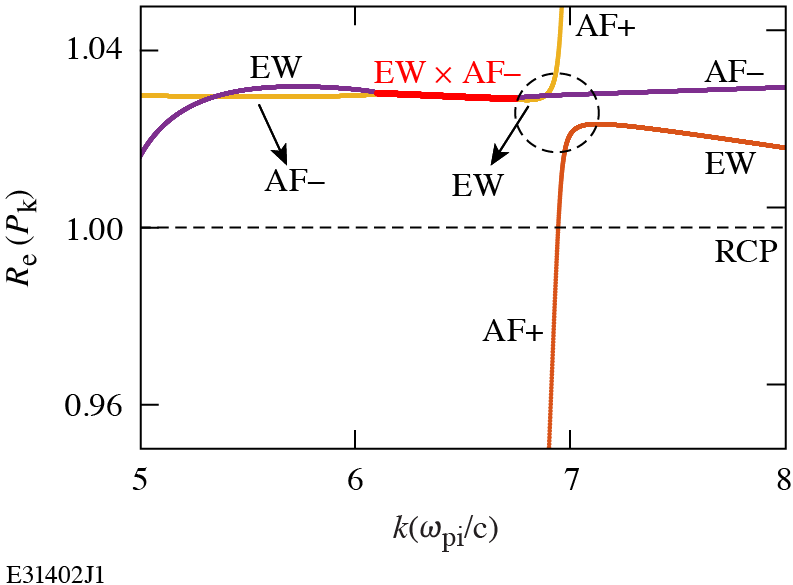}
    \caption{Polarization states of the modes at $\theta_k = 64^\circ$ near right-hand circular polarization (RCP), $P_k = +1$.}
    \label{fig:polarization}
\end{figure}

Figure~\ref{fig:polarization} shows the polarization states of the modes, calculated from the dispersion relation, in the $\theta_k = 64^\circ$ direction for the I: $\alpha = 0^\circ$ setup. An avoided crossing in their polarization states near $k\approx 7\omega_{pi}/c$ between EW and AF+ does not result in a resonance, but a mode transformation \cite{keppens2019fresh} instead [see inset in Fig. 3(b)].

\section{Ion-ion coupling MS$-$(i1)$\times$MS$-$(i2)}
\begin{figure}[H]
    \centering
    \includegraphics[width=0.7\linewidth]{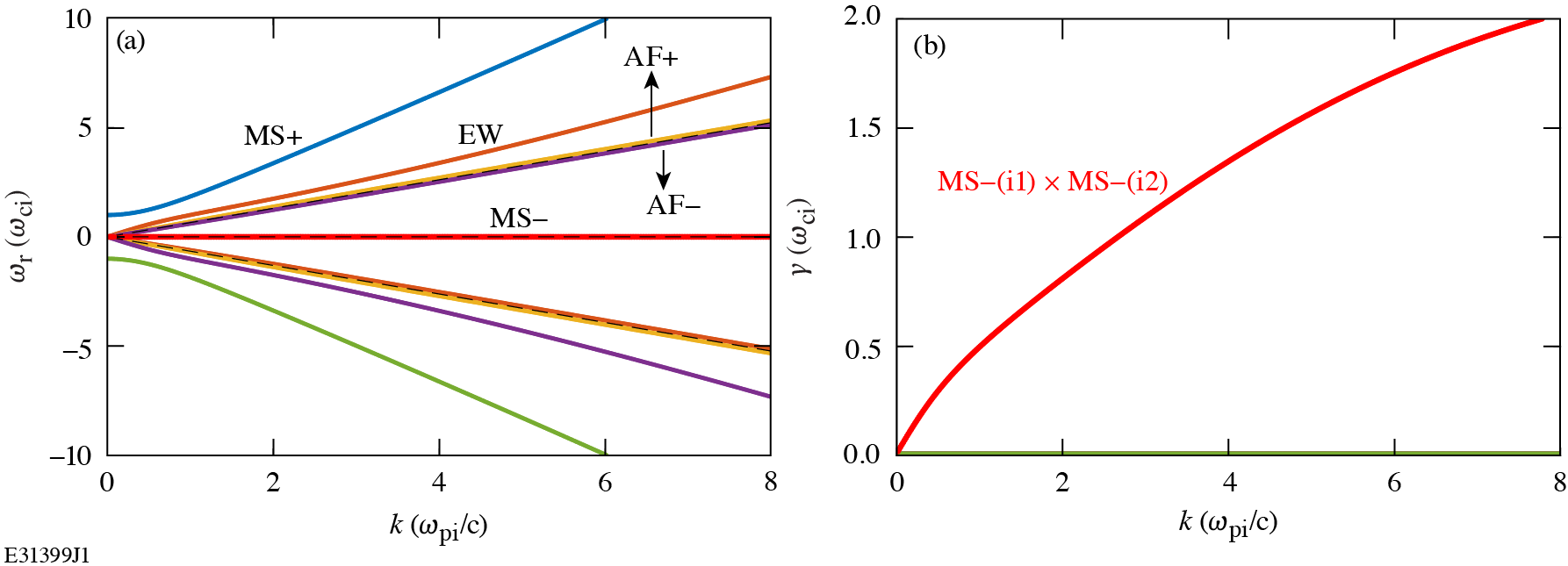}
    \caption{Multi-fluid dispersion relation at $\theta_k = 84^\circ$: (a) real frequency and (b) growth rate.}
    \label{fig:theta-84deg}
\end{figure}

Similar to Figs. 3(b) and (c), using the configuration for the I: $\alpha = 0^\circ$ simulations, Fig.~\ref{fig:theta-84deg} shows the multi-fluid dispersion relation at $\theta_k = 84^\circ$, marked by the white dotted line in Fig. 3(a). Unlike Figs. 3(b) and (c), here we show wave modes in both ion 1 and ion 2 species. 

In Fig.~\ref{fig:theta-84deg}(a), positively and negatively propagating waves are plotted in the upper ($\omega_r/k > 0$) and lower ($\omega_r/k < 0$) halves, respectively. Similar to the identification of the resonance between EW and ion MHD waves (EW$\times$MS$-$ and EW$\times$AF$-$ in Fig. 3), the unstable mode in Fig.~\ref{fig:theta-84deg} results from the resonance between two backward magnetosonic waves, MS$-$(i1)$\times$MS$-$(i2).

\section{Unmagnetized simulations}
\begin{figure}[H]
    \centering
    \includegraphics[width=0.4\linewidth]{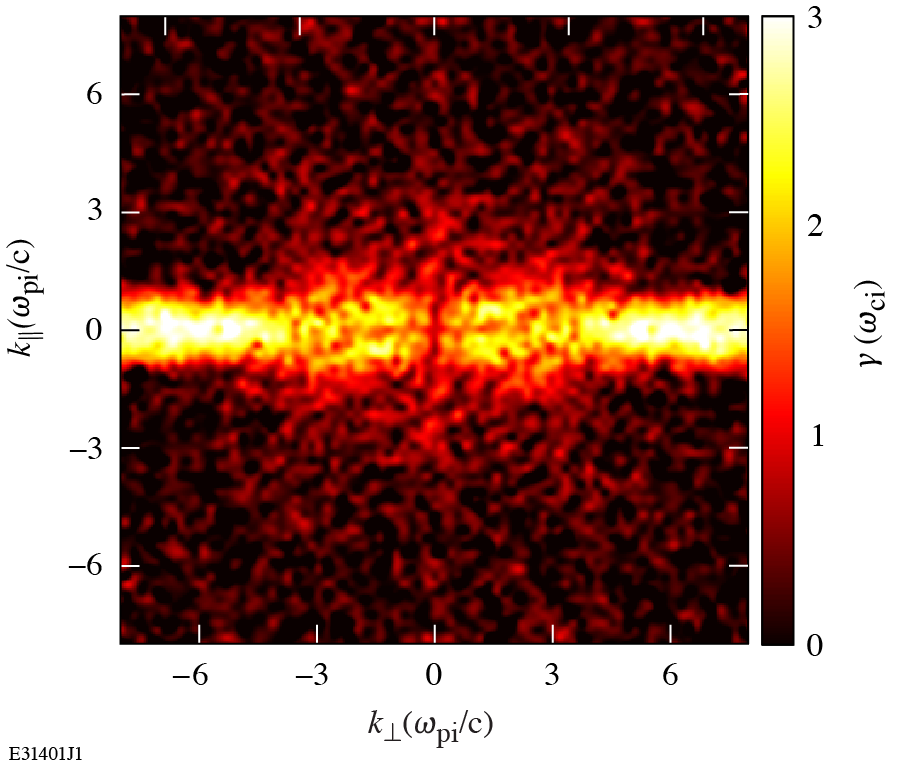}
    \caption{Growth rates inferred from the out-of-plane $B$-field for the I: $B$ = 0 simulations.}
    \label{fig:B=0}
\end{figure}

The importance of magnetization is examined by turning off the applied $B$-field. Figure~\ref{fig:B=0} shows the linear growth rates inferred from the out-of-plane $B$-field over 0 – $2\omega_{ci}^{-1}$ (same normalization as in the magnetized cases) for the I: $B$ = 0 simulations. Similar to the nearly-perpendicular magnetosonic mode [MS$-$(i1)$\times$MS$-$(i2)], the unstable acoustic waves from ion 1 and ion 2 species, respectively, produce a perpendicular mode in the spectrum. 

Compared to Fig. 3(a), the absence of a mode split at $k_{\parallel}$ = 0 in Fig.~\ref{fig:B=0} may be due to the difference between the magnetosonic and acoustic modes, because the MS mode has fast and slow modes in the nearly-perpendicular directions, while the acoustic mode is isotropic. Nevertheless, in the unmagnetized simulations, the purple dashed lines (I: $B$ = 0) in Fig. 1 indicate that the ion-ion acoustic (magnetosonic) mode does not drive an efficient ion-electron energy exchange.

\section{Isotropic electron heating}
\begin{figure}[H]
    \centering
    \includegraphics[width=0.4\linewidth]{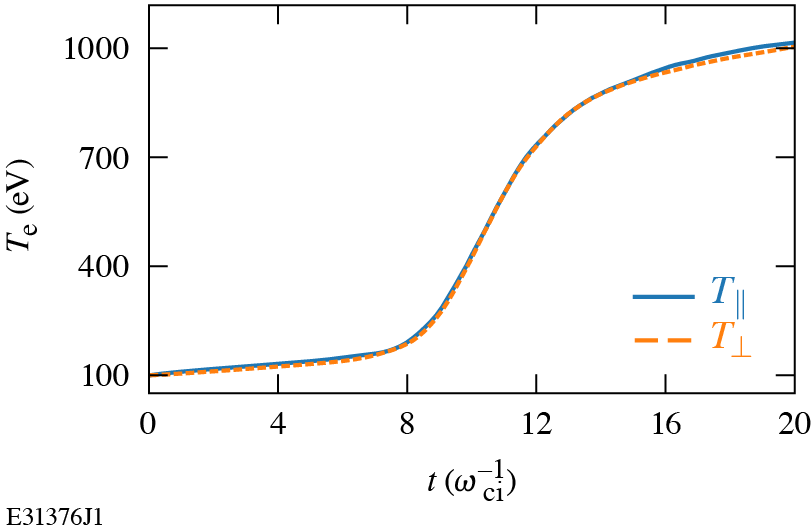}
    \caption{Time history of electron parallel and perpendicular temperatures.}
    \label{fig:e-heating}
\end{figure}

Supported by the instabilities whose growth rates are on the order of electron frequency, such as magnetic mirroring \cite{chen2015introduction} and anisotropy-driven instabilities \cite{gary2006linear,zhang2024collisionless}, a nearly isotropic electron heating on the ion timescale is observed, shown in Fig.~\ref{fig:e-heating}. 

\section{Realistic ion-electron mass ratio}
\begin{figure}[H]
    \centering
    \includegraphics[width=0.4\linewidth]{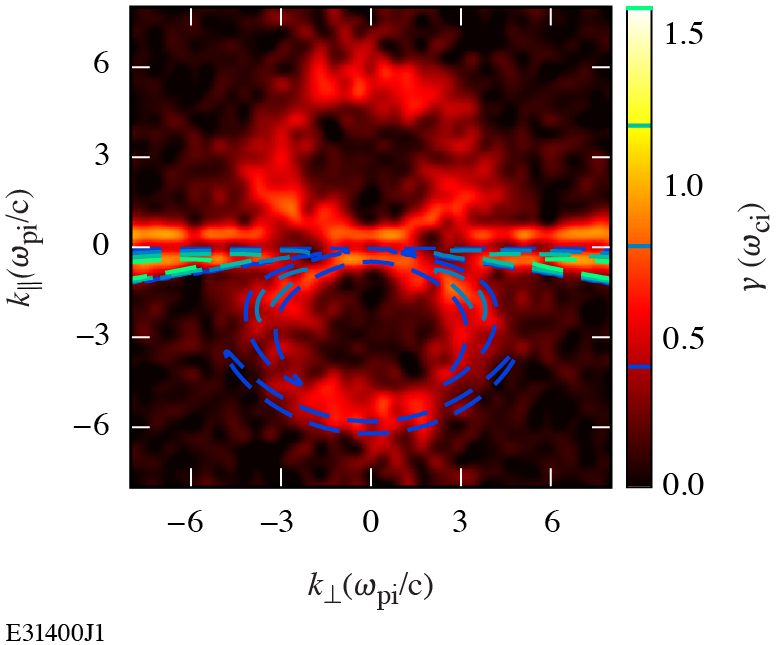}
    \caption{Growth rates inferred from the out-of-plane $B$-field for the simulations with a realistic ion-electron mass ratio of $m_i = 1836 m_e$. The contours in the lower half show the solutions from the multi-fluid dispersion relation.}
    \label{fig:mi-1836}
\end{figure}

The Letter utilizes a reduced ion-electron mass ratio of $m_i = 400 m_e$. For comparison, we ran simulations with a realistic mass ratio of $m_i = 1836 m_e$, using the same setup as the I: $\alpha = 0^\circ$ simulations: $\alpha = 0^\circ$, $n_{i1} = n_{i2}$, $T$ = 100 eV, and $M_{A,i1}$, $M_{A,i2}$ = $\pm6.25$. Figure~\ref{fig:mi-1836} shows the growth rate inferred from the out-of-plane $B$-field over 0 – $4\omega_{ci}^{-1}$. 

The growth rate shows a similar pattern to Fig. 3(a), where $m_i = 400 m_e$ was used. Therefore, we expect the resonant modes proposed in the Letter to work comparably and result in similar ion and electron heating in realistic cases.


\bibliographystyle{apsrev4-2}
\bibliography{Resonance}